\documentclass{article}

\usepackage{psfrag,epsfig,graphicx,graphics,xcolor}
\usepackage{wrapfig}
\usepackage{amsmath}
\usepackage{amssymb}
\newcommand{\bfr}{\begin{flushright}}
\newcommand{\efr}{\end{flushright}}

\def\slashchar#1{\setbox0=\hbox{$#1$}
   \dimen0=\wd0
   \setbox1=\hbox{/} \dimen1=\wd1
   \ifdim\dimen0>\dimen1
      \rlap{\hbox to \dimen0{\hfil/\hfil}}
      #1
   \else
      \rlap{\hbox to \dimen1{\hfil$#1$\hfil}}
      /
   \fi}

\def\bei{\begin{itemize}}
\def\ei{\end{itemize}}

\def\beeq{\begin{eqnarray}} 
\def\beqa{\begin{eqnarray}}
\def\bea{\begin{eqnarray}}

\def\eea{\end{eqnarray}}
\def\eqa{\end{eqnarray}}
\def\eeeq{\end{eqnarray}}

\def\eqar{\end{array}}
\def\beqar{\begin{array}}

\def\beas{\begin{eqnarray*}}
\def\beqas{\begin{eqnarray*}}

\def\eqas{\end{eqnarray*}}
\def\eeas{\end{eqnarray*}}

\def\beq{\begin{equation}} 
\def\be{\begin{equation}}

\def\ee{\end{equation}}
\def\eq{\end{equation}}
\def\eeq{\end{equation}}

\def\beqd{\begin{displaymath}}
\def\eeqd{\end{displaymath}}
\def\eqd{\end{displaymath}}

\def\beeq{\begin{eqnarray}} \def\eeeq{\end{eqnarray}}

\newcommand{\fin}{\end{document}}

\newcommand{\veck}{{\bf k}}

\newcommand{\veckone}{{\bf k}_1}
\newcommand{\vecktwo}{{\bf k}_2}
\newcommand{\veckj}{{\bf k}_{J}}

\newcommand{\veckjone}{{\bf k}_{J,1}}
\newcommand{\veckjtwo}{{\bf k}_{J,2}}

\newcommand{\deins}[1]{{\rm d}#1\,}
\newcommand{\dzwei}[1]{{\rm d}^2#1\,}
\newcommand{\dk}{\dzwei{\veck}}

\newcommand{\dkone}{\dzwei{\veckone}}
\newcommand{\dktwo}{\dzwei{\vecktwo}}

\newcommand{\dsigma}{\deins{\sigma}}
\newcommand{\dsigmahat}{\deins{{\hat\sigma}_{\rm{ab}}}}
\newcommand{\dnu}{\deins{\nu}}

\newcommand{\dx}{\deins{x}}
\newcommand{\dxone}{\deins{x_1}}
\newcommand{\dxtwo}{\deins{x_2}}
\newcommand{\dyjetone}{\deins{y_{J,1}}}
\newcommand{\dyjettwo}{\deins{y_{J,2}}}
\newcommand{\dphij}{\deins{\phi_{J}}}
\newcommand{\dphijone}{\deins{\phi_{J,1}}}
\newcommand{\dphijtwo}{\deins{\phi_{J,2}}}
\newcommand{\dtwojets}{{\rm d}|\veckjone|\,{\rm d}|\veckjtwo|\,\dyjetone \dyjettwo}

\newcommand{\shat}{{\hat s}}

\newcommand{\asbar}{{\bar{\alpha}}_s}

\newcommand{\avgcosn}{\langle \cos n \varphi \rangle}
\newcommand{\avgcos}{\langle \cos \varphi \rangle}
\newcommand{\avgcostwo}{\langle \cos 2 \varphi \rangle}
\newcommand{\avgcosthree}{\langle \cos 3 \varphi \rangle}

\begin{document}
\title{Can one use Mueller-Navelet jets at LHC as a clean test of QCD resummation effects at high energy?
\thanks{Presented at  the Low x workshop, May 30 - June 4 2013, Rehovot and
Eilat, Israel}
}

\author{
Bertrand Duclou\'e$^1$, Lech Szymanowski$^2$, Samuel Wallon$^{1,3}$\\
{\small $^1$LPT, Universit\'e Paris-Sud, CNRS, 91405, Orsay, France}\\
{\small $^2$National Centre for Nuclear Research (NCBJ), Warsaw, Poland}\\
{\small $^3$UPMC Univ. Paris 06, Facult\'e de physique,}\\
{\small 4 place Jussieu, 75252 Paris Cedex 05,
France
}
\smallskip\\
}

\date{\today
}

\maketitle
\begin{abstract}
The measurement of azimuthal correlations of
Mueller-Navelet jets is generally considered as a decisive test to reveal the effect of BFKL dynamics at hadron colliders. The first experimental study of these correlations at the LHC has been  recently performed by the CMS collaboration.
We show that the ratios of cosine moments of the azimuthal distribution
are successfully described within our next-to-leading logarithmic BFKL treatment. The whole set of CMS data for 
the azimuthal correlations can also be consistently described provided that one uses a larger renormalization/factorization scale than its natural value.

\end{abstract}

\section{Introduction}

The understanding of the high energy limit of QCD, in the so-called perturbative Regge limit, has been the subject of many studies. Many observables have been suggested to test these dynamics, based on inclusive~\cite{test_inclusive}, semi-inclusive~\cite{test_semi_inclusive} and exclusive processes~\cite{test_exclusive}.
In this limit, the smallness of the strong coupling $\alpha_s$ can be compensated by large logarithmic enhancements of the type $[\alpha_s\ln(s/|t|)]^n$ which have to be resummed, giving rise to the 
leading logarithmic (LL) Balitsky-Fadin-Kuraev-Lipatov (BFKL) Pomeron \cite{BFKL_LL}. Mueller and Navelet proposed to study the production of two jets with a large rapidity separation at hadron colliders \cite{Mueller:1986ey}. In a pure leading order collinear treatment these two jets would be emitted back to back, while a BFKL approach allows some emission between these jets which should lead to a larger cross section and lower angular correlation of the jets. We present results of a full next-to-leading logarithmic (NLL) analysis, in which the NLL corrections are included for the BFKL Green's function~\cite{BFKL_NLL} %~\cite{Fadin:1998py,Ciafaloni:1998gs} 
and the jet vertices~\cite{Bartels:vertex,Caporale:2011cc}.

In the following we will focus on the azimuthal correlations $\avgcosn$~\cite{DelDuca:Stirling} and ratios of these observables, as well as on the azimuthal distribution, at a center of mass energy $\sqrt{s}=7$ TeV, which have been measured recently at the LHC by the CMS collaboration~\cite{CMS-PAS-FSQ-12-002}. We make some comparison of our results~\cite{Ducloue:2013hia} with these data and investigate the dependency on the various scales, including the renormalization scale.

\section{Basic formulas}

\begin{figure}[htbp]
\centering
\includegraphics[height=8cm]{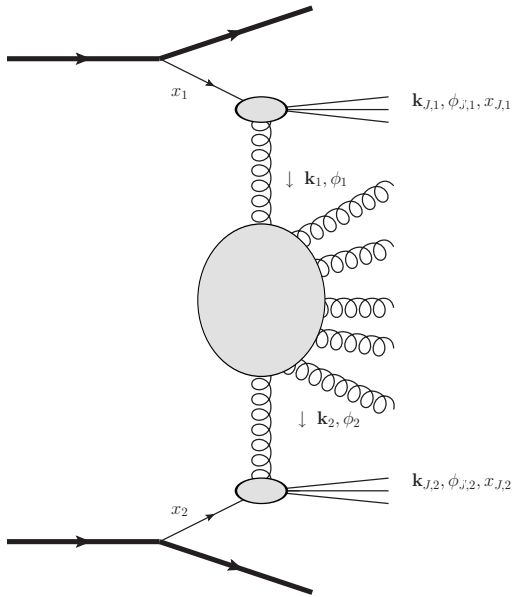}
\caption{Kinematics of the process}
\label{Fig:kinematics}
\end{figure}

Let us consider, as shown on Fig.~\ref{Fig:kinematics}, two hadrons colliding at a center of mass energy $\sqrt{s}$. Relying on the collinear factorization, the differential cross section reads
\beqa
  &&\frac{\dsigma}{\dtwojets} \nonumber\\
  &&= \sum_{{\rm a},{\rm b}} \int_0^1 \dxone \int_0^1 \dxtwo f_{\rm a}(x_1) f_{\rm b}(x_2) \frac{\dsigmahat}{\dtwojets},
\eqa
where $\veckjone$, $\veckjtwo$ are the transverse momenta of the jets, $y_{J,1}$ and $y_{J,2}$ their rapidities and $f_{\rm a}$  $(f_{\rm  b})$ are the parton distribution functions (PDFs) of a parton a (b) in the according proton. In this expression, the partonic cross section can be expressed as
\beqa
 && \frac{\dsigmahat}{\dtwojets} \nonumber\\
  &&= \int \dphijone\dphijtwo\int\dkone\dktwo V_{\rm a}(-\veckone,x_1)\,G(\veckone,\vecktwo,\shat)\,V_{\rm b}(\vecktwo,x_2),\label{eq:bfklpartonic}
\eqa
where $\phi_{J,1}$ and $\phi_{J,2}$ are the azimuthal angles of the jets, $V_a$ $(V_b)$ is the jet vertex initiated by the parton a (b) and $G$ is the BFKL Green's function which depends on $\shat=x_1 x_2 s$. For further use, 
it is convenient to introduce the coefficients $\mathcal{C}_n$, defined as
\begin{equation}
  \mathcal{C}_n = (4-3\delta_{n,0}) \int \dnu C_{n,\nu}(|\veckjone|,x_{J,1})C^*_{n,\nu}(|\veckjtwo|,x_{J,2}) \left( \frac{\shat}{s_0} \right)^{\omega(n,\nu)}\,,
  \label{Cn}
\end{equation}
such that
\begin{equation}
\label{def:dsigma}
\frac{\dsigma}{\dtwojets} = \mathcal{C}_0 \,,
\end{equation}
and
\begin{equation} 
\label{def:decor}
  \langle\cos(n\varphi)\rangle \equiv \langle\cos\big(n(\phi_{J,1}-\phi_{J,2}-\pi)\big)\rangle = \frac{\mathcal{C}_n}{\mathcal{C}_0} \,.
\end{equation}
In Eq.~(\ref{Cn}), $C_{n,\nu}$ is defined as
\begin{equation}
   C_{n,\nu}(|\veckj|,x_{J})= \int\dphij\dk \dx f(x) V(\veck,x) E_{n,\nu}(\veck) \cos(n\phi_J)\,,
  \label{Cnnu}
\end{equation}
with $E_{n,\nu}$ being the LL BFKL eigenfunctions 
\begin{equation}
  E_{n,\nu}(\veckone) = \frac{1}{\pi\sqrt{2}}\left(\veckone^2\right)^{i\nu-\frac{1}{2}}e^{in\phi_1}\,.
\label{def:eigenfunction}
\end{equation}
At LL accuracy, the eigenvalue $\omega(n,\nu)$ of the BFKL kernel is
\begin{equation}
  \omega(n,\nu) = \asbar \chi_0\left(|n|,\frac{1}{2}+i\nu\right)\,,
\end{equation}
with  
\begin{equation}
 \chi_0(n,\gamma) = 2\Psi(1)-\Psi\left(\gamma+\frac{n}{2}\right)-\Psi\left(1-\gamma+\frac{n}{2}\right)\,,
\end{equation}
where $\asbar = \alpha_s N_c/\pi$, $\Psi(z) = \Gamma'(z)/\Gamma(z)$
and the vertex is
\begin{equation}
  V_{\rm a}(\veck,x)=V_{\rm a}^{(0)}(\veck,x) = \frac{\alpha_s}{\sqrt{2}}\frac{C_{A/F}}{\veck^2} \delta\left(1-\frac{x_J}{x}\right)|\veckj|\delta^{(2)}(\veck-\veckj)\,,
\end{equation}
with $C_A$ for ${\rm a}={\rm g}$ and $C_F$ for ${\rm a}={\rm q}$. At NLL, despite the fact that the $E_{n,\nu}$
are not eigenfunctions of the kernel anymore due to conformal
invariance breaking terms, it is still possible to use them, the price to pay being an explicit dependency on $\veckjone$ and $\veckjtwo$~\cite{Kotikov:2000pm-Kotikov:2002ab-Ivanov:2005gn-Vera:2006un,Vera:2007kn,Schwennsen:2007hs}:
\beqa
 && \omega(n,\nu) = \asbar \chi_0\left(|n|,\frac{1}{2}+i\nu\right) \nonumber \\
  &&+ \asbar^2 \left[ \chi_1\left(|n|,\frac{1}{2}+i\nu\right)-\frac{\pi b_0}{N_c}\chi_0\left(|n|,\frac{1}{2}+i\nu\right) \ln\frac{|\veckjone|\cdot|\veckjtwo|}{\mu_R^2} \right]\,,
\eqa
with $b_0=(33 - 2 \,N_f)/(12 \pi)$
and $V_{\rm a}(\veck,x) = V^{(0)}_{\rm a}(\veck,x) + \alpha_s V^{(1)}_{\rm a}(\veck,x)$. The expression for the NLL corrections to the Green's function resulting in $\chi_1$ can be found in Eq.~(2.17) of Ref.~\cite{Ducloue:2013hia}. The expressions of the NLL corrections to the jet vertices are quite lengthy and will not be reproduced here. They can be found in Ref.~\cite{Colferai:2010wu}, as extracted from Refs.~\cite{Bartels:vertex} 
after correcting a few misprints.  They have been recently reobtained in Ref.~\cite{Caporale:2011cc}. In the limit of small cone jets, they have been computed in Ref.~\cite{Ivanov:2012ms} and applied to phenomenology in Refs.~\cite{Caporale:2012ih,Caporale:2013uva}.
Here we use the cone algorithm with a size of $R_{\rm cone}=0.5$. Note, however, that using the $k_t$ or 
the anti-$k_t$ algorithm leads to negligible changes in our predictions. Our calculation depends on the renormalization scale $\mu_R$, the factorization scale $\mu_F$ and the energy scale $\sqrt{s_0}$. In the following we set $\mu_R=\mu_F\equiv\mu$. We choose the ``natural'' value $\sqrt{|\veckjone|\cdot |\veckjtwo|}$ for $\mu$ and $\sqrt{s_0}$, and vary these scales by a factor of 2 to estimate the scale uncertainty of our calculation. We use the MSTW 2008 PDFs \cite{Martin:2009iq} and a two-loop running coupling. We also include collinear improvement to the Green's function as was suggested in Refs.~\cite{Salam:1998tj-Ciafaloni:1998iv-Ciafaloni:1999yw-Ciafaloni:2003rd} and extended for $n \neq 0$ in Refs.~\cite{Vera:2007kn,Schwennsen:2007hs,Marquet:2007xx}.

\section{Results: asymmetric configuration}

In \cite{Ducloue:2013hia} we performed a detailed study of several BFKL scenarios, from a pure LL approximation (LL Green's function and leading order jet vertex) to a full NLL calculation (NLL Green's function and NLL jet vertex). The main conclusion is that there is a dramatic effect when passing from a mixed treatment, which combines LL vertices with NLL Green's function, to a full NLL approach, of magnitude similar to the one when passing from 
a pure LL result to the mixed treatment: the inclusion of NLL corrections inside the jets vertices is therefore essential. 
Note that the effect of collinear improvement in the NLL Green's function is very small when using the NLL jet vertices.
The inclusion of the whole set of NLL corrections leads to decorrelation effects which are much smaller than expected. Still, there remains a sizable difference between the full NLL approach and the fixed order predictions at next-to-leading order (NLO)
for the ratios $\avgcostwo / \avgcos$ and $\avgcosthree / \avgcostwo$, when taking into account theoretical uncertainties.

\begin{figure}[htbp]
  \def\sca{.6}
  \psfrag{central}[l][l][0.5]{full NLL}
  \psfrag{muchange_0.5}[l][l][\sca]{\footnotesize $\mu \to \mu/2$}
  \psfrag{muchange_2.0}[l][l][\sca]{\footnotesize $\mu \to2 \mu$}
  \psfrag{s0change_0.5}[l][l][\sca]{\footnotesize $\sqrt{s_0} \to \sqrt{s_0}/2$}
  \psfrag{s0change_2.0}[l][l][\sca]{\footnotesize $\sqrt{s_0} \to 2 \sqrt{s_0}$}
  \psfrag{Dijet}[l][l][0.5]{fixed order NLO}
  \psfrag{Y}{\scalebox{0.8}{\hspace{-0.1cm}$Y$}}
  \begin{minipage}{0.49\textwidth}
  \psfrag{cos}{\scalebox{0.8}{$\langle \cos 2 \varphi\rangle / \langle \cos \varphi\rangle$}}
  \hspace{-.3cm}  \includegraphics[width=6.3cm]{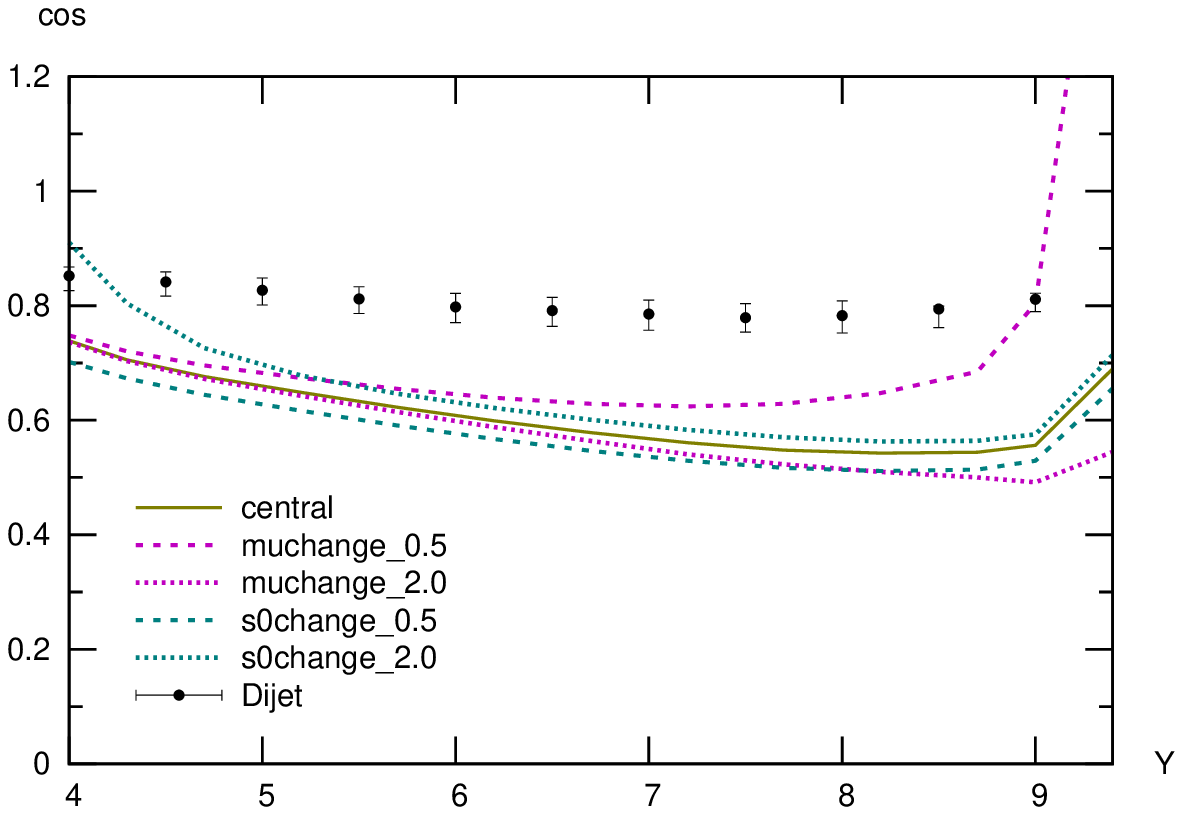}
  \end{minipage}
  \begin{minipage}{0.49\textwidth}
  \psfrag{cos}{\scalebox{0.8}{$\langle \cos 3 \varphi\rangle / \langle \cos 2\varphi\rangle$}}
  \hspace{-.3cm}  \includegraphics[width=6.3cm]{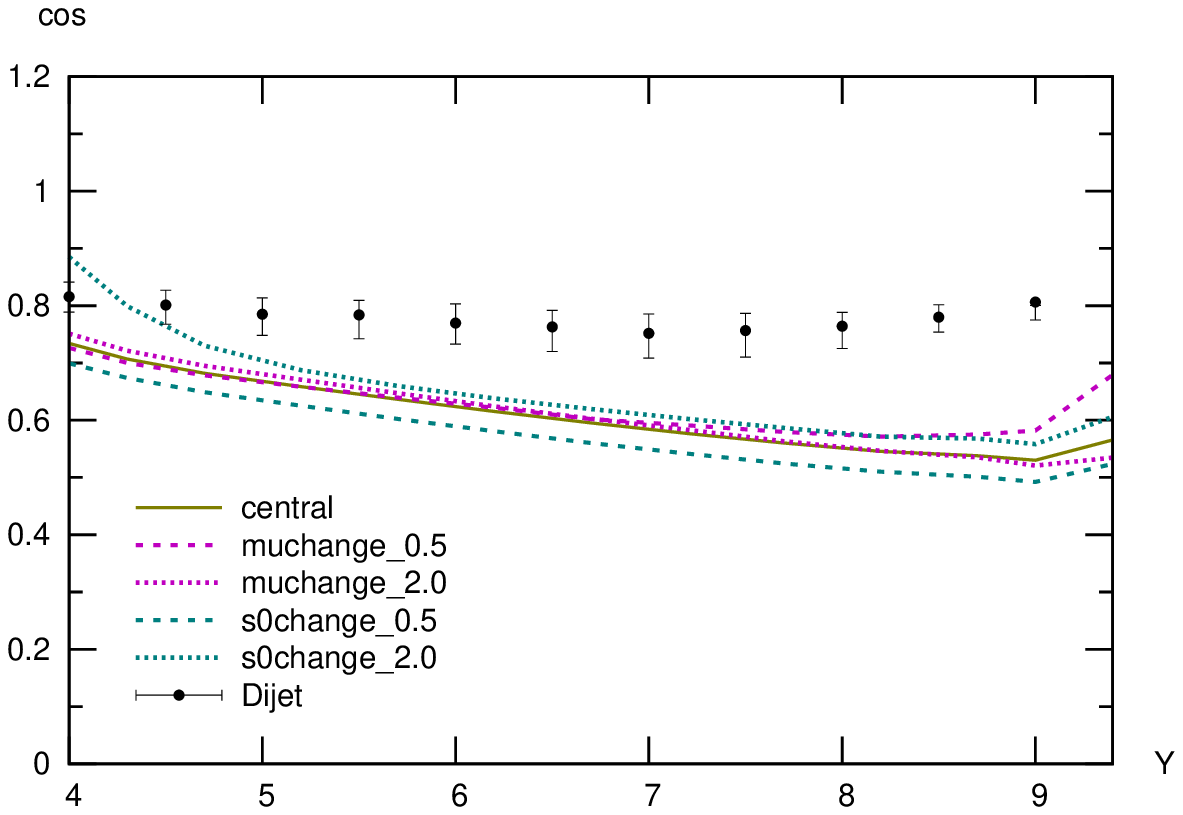}
  \end{minipage}
   \caption{Comparison of the full NLL BFKL calculation including the scale uncertainty with \textsc{Dijet} predictions, using asymmetric cuts defined in Eq.~(\protect\ref{asym-cuts}). Left: value of $\avgcostwo / \avgcos$ as a function of the rapidity separation $Y$.  Right: value of $\avgcosthree / \avgcostwo$ as a function of the rapidity separation $Y$.}
\label{Fig:cos2cos_cos3cos2_asym}
\end{figure}

To study the need for high-energy resummation with respect to fixed order treatments, one should however pay attention to the fact that
these fixed order calculations have instabilities when the lower cuts on the transverse momenta of the jets are identical. Thus, at the moment, it is not possible to make a direct comparison between these two predictions in an asymmetric configuration and  the data of Ref.~\cite{CMS-PAS-FSQ-12-002}, which have been extracted in a symmetric configuration. Still, one can compare our BFKL calculation with the fixed order NLO code \textsc{Dijet}~\cite{Aurenche:2008dn} in an asymmetric configuration with the following cuts, which could be implemented by CMS:
\begin{eqnarray}
  35\,{\rm GeV} < &|\veckjone|, |\veckjtwo| & < 60 \,{\rm GeV} \,, \nonumber\\
  50\,{\rm GeV} < &{\rm Max}(|\veckjone|, |\veckjtwo|)\,, \nonumber\\
 0 < &y_1, \, y_2& < 4.7\,.
 \label{asym-cuts}
\end{eqnarray}
This is illustrated for the ratios $\avgcostwo / \avgcos$ and $\avgcosthree / \avgcostwo$ in Fig.~\ref{Fig:cos2cos_cos3cos2_asym}. In this figure, we show the variation of our NLL result when varying $\mu$ and $\sqrt{s_0}$ by a factor of 2 and compare it with the \textsc{Dijet} prediction. Here the fixed order NLO calculation is significantly above the full NLL BFKL calculation. These observables are quite stable with respect to the scales so that the difference between NLL BFKL and fixed order NLO does not vanish when we take into account the scale uncertainty.

\section{Results: symmetric configuration}

In this section, we
compare our results with the data recently obtained by CMS~\cite{CMS-PAS-FSQ-12-002}, for a symmetric configuration (identical lower cut for the transverse momenta of the jets) and with cuts 
\begin{eqnarray}
 35\,{\rm GeV} < &|\veckjone|,|\veckjtwo|& < 60 \,{\rm GeV} \,, \nonumber\\
 0 < &y_1, \, y_2& < 4.7\,.
 \label{sym-cuts}
\end{eqnarray}
These are the cuts used by CMS in Ref.~\cite{CMS-PAS-FSQ-12-002}, with the exception that for numerical reasons we have to set an upper cut on the transverse momenta of the jets. We have checked that our results do not depend strongly on the value of this cut as the cross section is strongly peaked near the minimum value allowed for $\veckjone$ and $\veckjtwo$. This enables us to compare our predictions with LHC data.

\begin{figure}[htbp]
  \def\sca{.6}
  \psfrag{Y}{\scalebox{0.8}{\hspace{-0.1cm}$Y$}}
  \begin{minipage}{0.49\textwidth}
     \psfrag{central}[l][l][0.5]{full NLL}
     \psfrag{muchange_0.5}[l][l][\sca]{\footnotesize $\mu \to \mu/2$}
     \psfrag{muchange_2.0}[l][l][\sca]{\footnotesize $\mu \to2 \mu$}
     \psfrag{s0change_0.5}[l][l][\sca]{\footnotesize $\sqrt{s_0} \to \sqrt{s_0}/2$}
     \psfrag{s0change_2.0}[l][l][\sca]{\footnotesize $\sqrt{s_0} \to 2 \sqrt{s_0}$}
     \psfrag{CMS}[l][l][0.5]{CMS data}
     \psfrag{cos}{\scalebox{0.8}{$\langle \cos \varphi\rangle$}}
     \hspace{-.2cm}  \includegraphics[width=6.2cm]{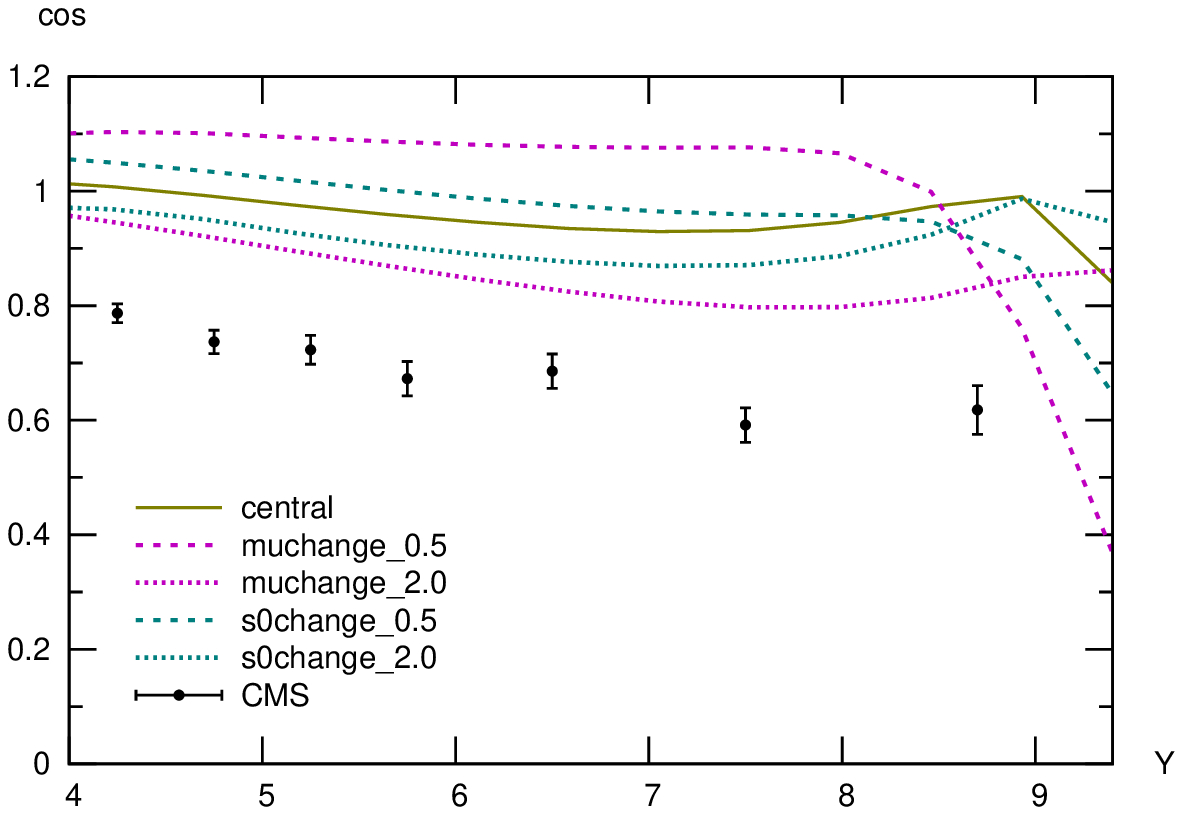}
  \end{minipage}
  \begin{minipage}{0.49\textwidth}
    \psfrag{central}[r][r][0.5]{full NLL}
    \psfrag{muchange_0.5}[r][r][\sca]{\footnotesize $\mu \to \mu/2$}
    \psfrag{muchange_2.0}[r][r][\sca]{\footnotesize $\mu \to2 \mu$}
    \psfrag{s0change_0.5}[r][r][\sca]{\footnotesize $\sqrt{s_0} \to \sqrt{s_0}/2$}
    \psfrag{s0change_2.0}[r][r][\sca]{\footnotesize $\sqrt{s_0} \to 2 \sqrt{s_0}$}
    \psfrag{CMS}[r][r][0.5]{CMS data}
    \psfrag{cos}{\scalebox{0.8}{$\langle \cos 2 \varphi\rangle$}}
    \hspace{-.2cm}  \includegraphics[width=6.2cm]{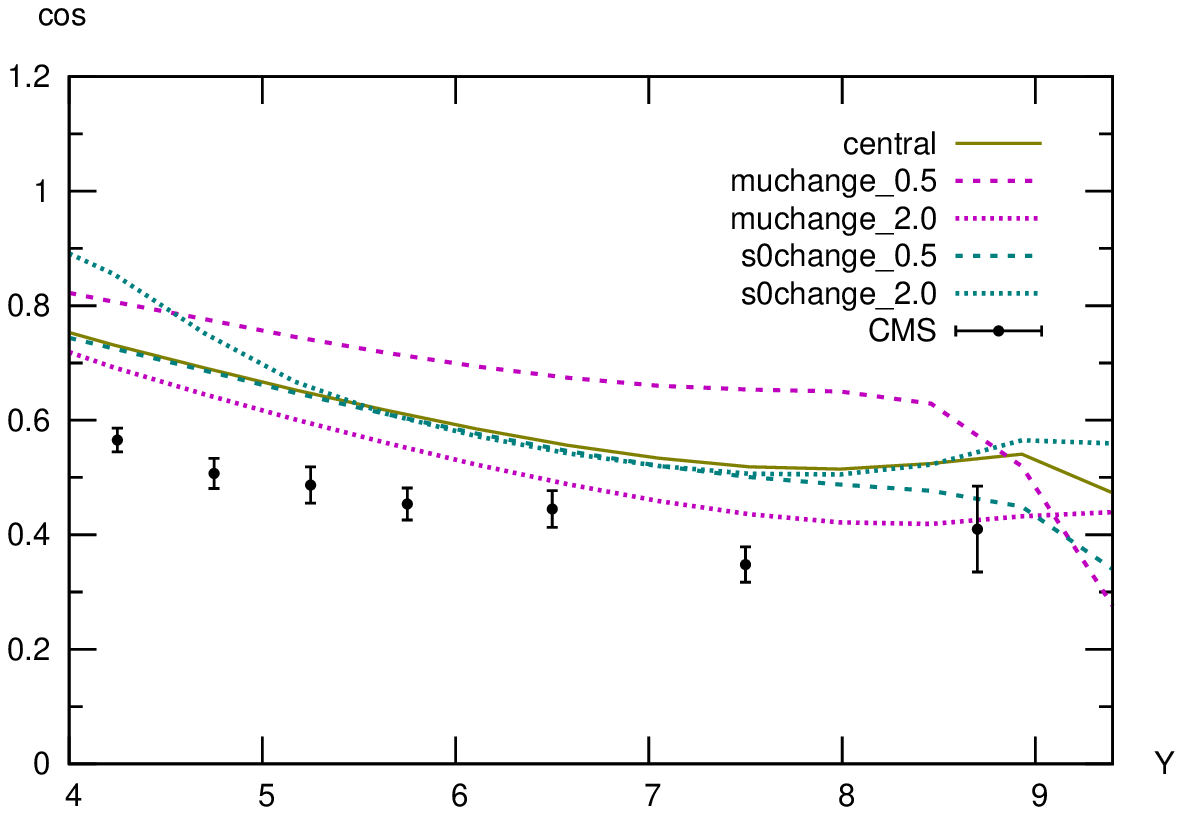}
  \end{minipage}
  \caption{Comparison of the full NLL BFKL calculation including the scale uncertainty with CMS data, using symmetric cuts defined in Eq.~(\protect\ref{sym-cuts}). Left: value of $\avgcos$ as a function of the rapidity separation $Y$. Right: value of $\avgcostwo$ as a function of the rapidity separation $Y$.}
  \label{Fig:cos_sym-cos2_sym}
\end{figure}

\begin{figure}[htbp]
  \def\sca{.6}
  \psfrag{central}[r][r][0.5]{full NLL}
  \psfrag{muchange_0.5}[r][r][\sca]{\footnotesize $\mu \to \mu/2$}
  \psfrag{muchange_2.0}[r][r][\sca]{\footnotesize $\mu \to2 \mu$}
  \psfrag{s0change_0.5}[r][r][\sca]{\footnotesize $\sqrt{s_0} \to \sqrt{s_0}/2$}
  \psfrag{s0change_2.0}[r][r][\sca]{\footnotesize $\sqrt{s_0} \to 2 \sqrt{s_0}$}
  \psfrag{CMS}[r][r][0.5]{CMS data}
  \psfrag{cos}{\scalebox{0.8}{$\langle \cos 3\varphi\rangle$}}
  \psfrag{Y}{\scalebox{0.8}{\hspace{-0.1cm}$Y$}}
  \begin{minipage}{0.49\textwidth}
     \hspace{-.2cm} \includegraphics[width=6.2cm]{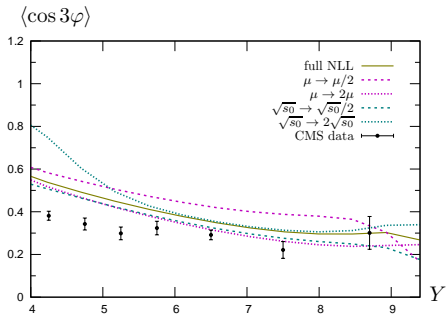}
  \end{minipage}
  \caption{Comparison of the full NLL BFKL calculation including the scale uncertainty with CMS data, using symmetric cuts defined in Eq.~(\protect\ref{sym-cuts}), for $\avgcosthree$ as a function of the rapidity separation $Y$.}
  \label{Fig:cos3_sym}
\end{figure}

We begin our analysis with the azimuthal correlations $\avgcosn$. In Figs.~\ref{Fig:cos_sym-cos2_sym} and~\ref{Fig:cos3_sym} we show the variation of $\avgcos$, $\avgcostwo$ and $\avgcosthree$ with respect to the rapidity separation between the two jets $Y$
within a full NLL framework. We display the theoretical uncertainty obtained when varying $\mu$ and $s_0$ by a factor of 2 and we compare these predictions with CMS data (black dots with error bars). We see that NLL BFKL predicts a larger correlation than seen in the data, but these observables are strongly dependent on the value of the scales.

\begin{figure}[htbp]
  \def\sca{.6}
  \psfrag{central}[l][l][0.5]{full NLL}
  \psfrag{muchange_0.5}[l][l][\sca]{\footnotesize $\mu \to \mu/2$}
  \psfrag{muchange_2.0}[l][l][\sca]{\footnotesize $\mu \to2 \mu$}
  \psfrag{s0change_0.5}[l][l][\sca]{\footnotesize $\sqrt{s_0} \to \sqrt{s_0}/2$}
  \psfrag{s0change_2.0}[l][l][\sca]{\footnotesize $\sqrt{s_0} \to 2 \sqrt{s_0}$}
  \psfrag{CMS}[l][l][0.5]{CMS data}
  \psfrag{Y}{\scalebox{0.8}{\hspace{-0.1cm}$Y$}}
  \begin{minipage}{0.49\textwidth}
      \psfrag{cos}{\scalebox{0.8}{$\langle \cos 2 \varphi\rangle / \langle \cos \varphi\rangle$}}
      \hspace{-.2cm}  \includegraphics[width=6.2cm]{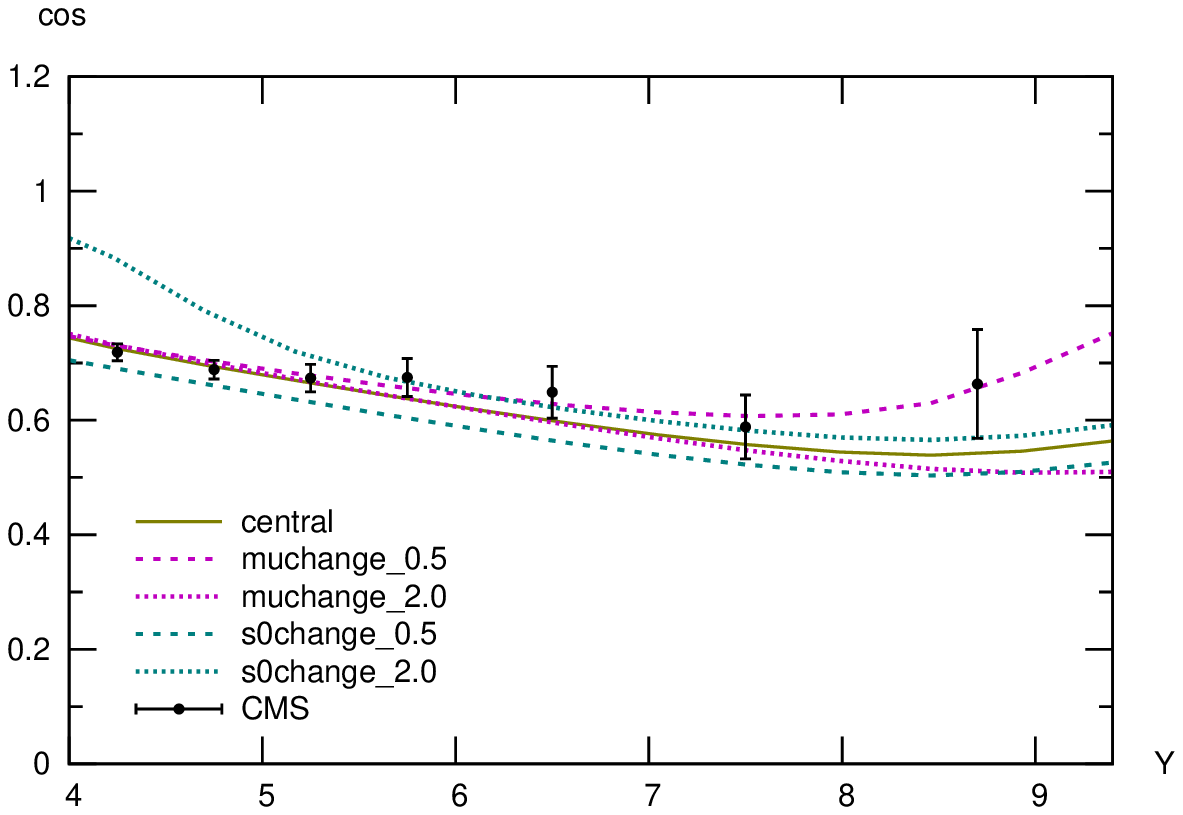}
  \end{minipage}
  \begin{minipage}{0.49\textwidth}
      \psfrag{cos}{\scalebox{0.8}{$\langle \cos 3 \varphi\rangle / \langle \cos 2\varphi\rangle$}}
      \hspace{-.2cm}  \includegraphics[width=6.2cm]{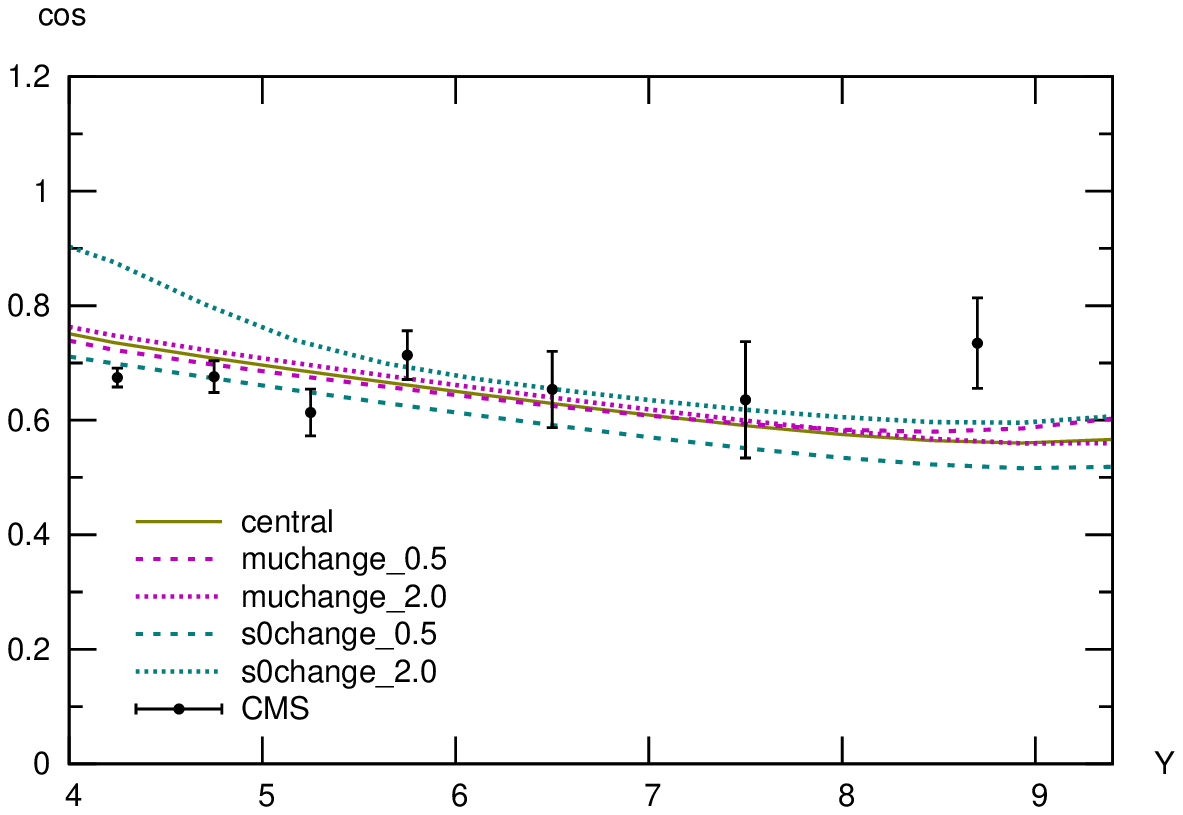}
  \end{minipage}
  \caption{Comparison of the full NLL BFKL calculation including the scale uncertainty with CMS data, using symmetric cuts defined in Eq.~(\protect\ref{sym-cuts}). Left: value of $\avgcostwo / \avgcos$ as a function of the rapidity separation $Y$. Right: value of $\avgcosthree / \avgcostwo$ as a function of the rapidity separation $Y$.}
  \label{Fig:cos2cos_cos3cos2_sym}
\end{figure}

The extraction of ratios of the previously mentioned observables was also performed in~\cite{CMS-PAS-FSQ-12-002}. In Fig.~\ref{Fig:cos2cos_cos3cos2_sym} we show results for $\avgcostwo / \avgcos$ and $\avgcosthree / \avgcostwo$. The effect of NLL corrections to the vertices is important (see Refs.~\cite{Ducloue:2013hia,Ducloue:2013tpa} for detailed comparisons), but these observables are more stable with respect to  $\mu$ and $s_0$ than the previous ones. The agreement with the data is very good over a large $Y$ range.

\begin{figure}[htbp]
  \psfrag{Y}{\scalebox{0.8}{\hspace{-0.1cm}$Y$}}
  \def\scale8{1}
  \begin{minipage}{0.49\textwidth}
    \psfrag{central}[l][l][.65]{\footnotesize $\mu=\sqrt{|\veckjone|\cdot |\veckjtwo|}$}
    \psfrag{muchange8s0change1}[l][l][.65]{\raisebox{-.2cm}{\footnotesize $\mu=8\sqrt{|\veckjone|\cdot |\veckjtwo|}$}}
    \psfrag{CMS}[l][l][.65]{\raisebox{-.3cm}{\footnotesize CMS data}}
    \psfrag{cos}{\scalebox{0.8}{$\langle \cos \varphi\rangle$}}
    \hspace{-.2cm}  \scalebox{\scale8}{\includegraphics[width=6.1cm]{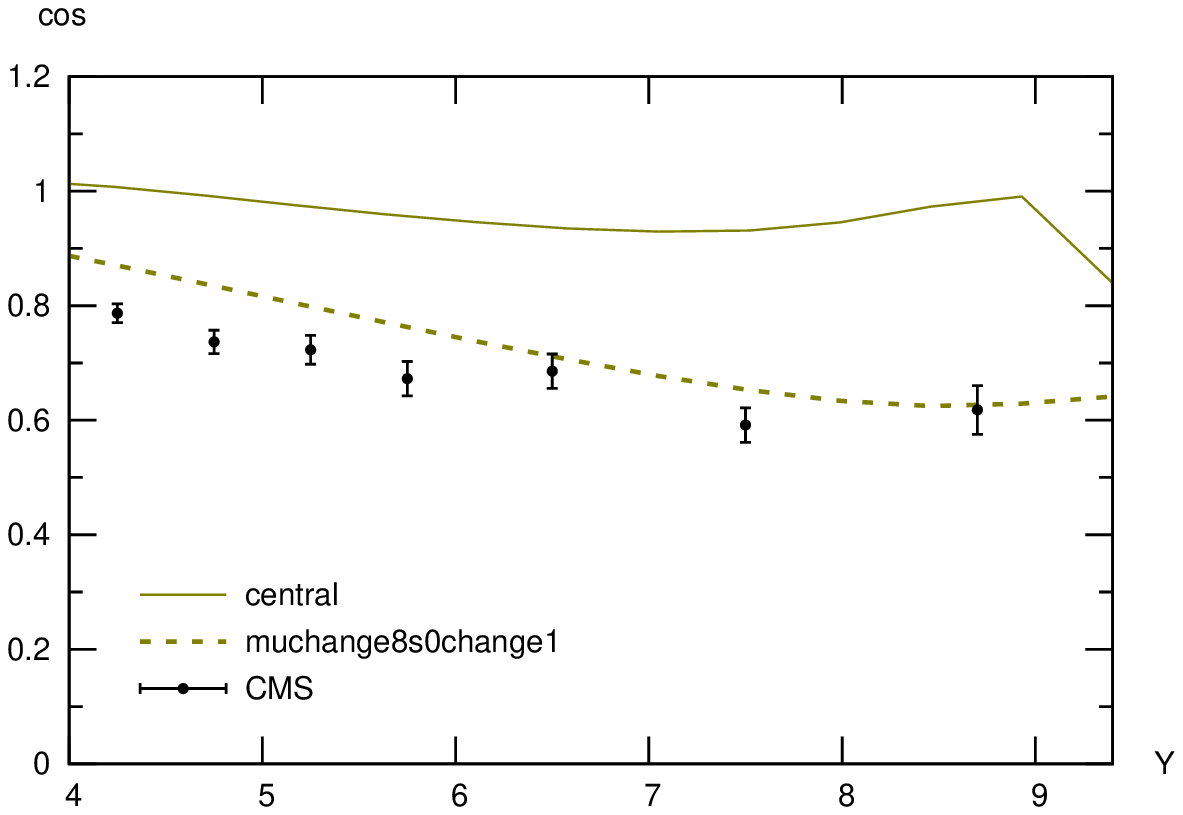}} 
  \end{minipage}
  \begin{minipage}{0.49\textwidth}
    \psfrag{central}[r][r][.65]{\footnotesize $\mu=\sqrt{|\veckjone|\cdot |\veckjtwo|}$}
    \psfrag{muchange8s0change1}[r][r][.65]{\raisebox{-.2cm}{\footnotesize $\mu=8\sqrt{|\veckjone|\cdot |\veckjtwo|}$}}
    \psfrag{CMS}[r][r][.65]{\raisebox{-.3cm}{ \footnotesize CMS data}}
    \psfrag{cos}{\scalebox{0.8}{$\langle \cos 2\varphi\rangle$}}
    \hspace{-.1cm}  \scalebox{\scale8}{\includegraphics[width=6.1cm]{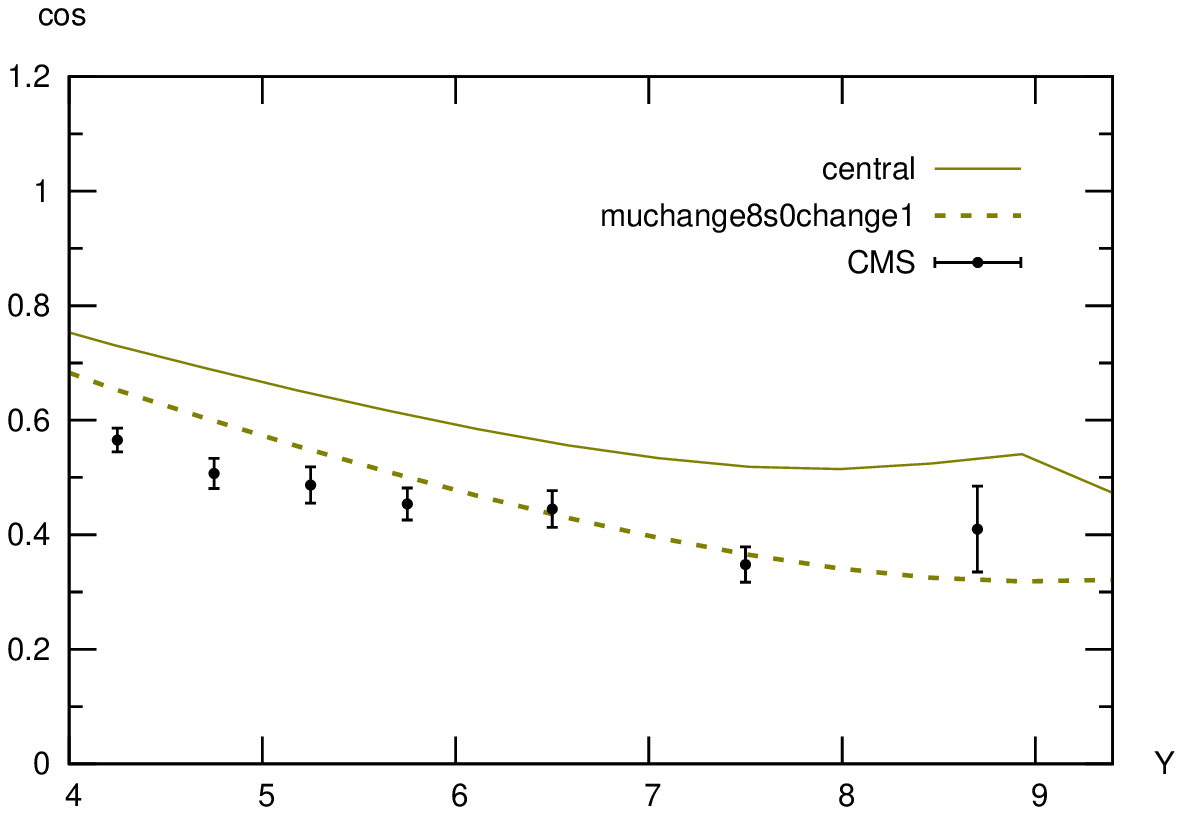}}
  \end{minipage}
  \caption{Comparison of the full NLL BFKL calculation with either a ``natural'' scale $\mu=\sqrt{|\veckjone|\cdot |\veckjtwo|}\,$
  or a large scale $\mu=8\sqrt{|\veckjone|\cdot |\veckjtwo|}\,$ and the CMS data, using symmetric cuts defined in (\protect\ref{sym-cuts}). Left: value of $\avgcos$ as a function of the rapidity separation $Y$. Right: value of $\avgcostwo$ as a function of the rapidity separation $Y$.}
\label{Fig:cos-cos2_scale8}
\end{figure}

Our code allows us to perform a complete study of the azimuthal distribution~\cite{Ducloue:2012bm},
which is most directly accessible in experiments, while more difficult to evaluate numerically
 than the individual moments discussed above. It is  defined as
\begin{equation}
  \frac{1}{{\sigma}}\frac{d{\sigma}}{d \varphi}
  ~=~ \frac{1}{2\pi}
  \left\{1+2 \sum_{n=1}^\infty \cos{\left(n \varphi\right)}
  \left<\cos{\left( n \varphi \right)}\right>\right\}\,.
  \label{dist-ang}
\end{equation}
If one tries to compare the full NLL BFKL prediction with CMS data, we fail to describe the distribution for large values of $\varphi$, corresponding to nearside jets configurations, when taking a ``natural'' scale $\mu=\sqrt{|\veckjone|\cdot |\veckjtwo|}$. We have seen also in Figs.~\ref{Fig:cos_sym-cos2_sym} and~\ref{Fig:cos3_sym} that the data is better described if we use larger values of $\mu$. Indeed, if we let the renormalization/factorization scale vary by more than a factor of 2, one can see that the whole set of CMS data can be very well described within the full NLL BFKL approach with a scale of the order of $\mu \sim 8 \sqrt{|\veckjone|\cdot |\veckjtwo|}\,.$ This is illustrated in Figs.~\ref{Fig:cos-cos2_scale8}, \ref{Fig:cos3-cos2cos_scale8} and~\ref{Fig:cos3cos2_dist_scale8}.
%
%
%%%%%%%%%%%%%%%%  cos and cos2 scale 8 muF   %%%%%%
%
%
%
%%%%%%%%%%%%  cos3 and cos2/cos scale 8 muF  %%%%%% 
%
%
\def\scale8{1}
\begin{figure}[htbp]
  \psfrag{Y}{\scalebox{0.8}{\hspace{-0.1cm}$Y$}}
  \begin{minipage}{0.49\textwidth}
    \psfrag{central}[r][r][.65]{\footnotesize $\mu=\sqrt{|\veckjone|\cdot |\veckjtwo|}$}
    \psfrag{muchange8s0change1}[r][r][.65]{\raisebox{-.2cm}{\footnotesize $\mu=8\sqrt{|\veckjone|\cdot |\veckjtwo|}$}}
    \psfrag{CMS}[r][r][.65]{\raisebox{-.3cm}{\footnotesize CMS data}}
    \psfrag{cos}{\scalebox{0.8}{$\langle \cos 3\varphi\rangle$}}
    \hspace{-.2cm}\scalebox{\scale8}{\includegraphics[width=6.1cm]{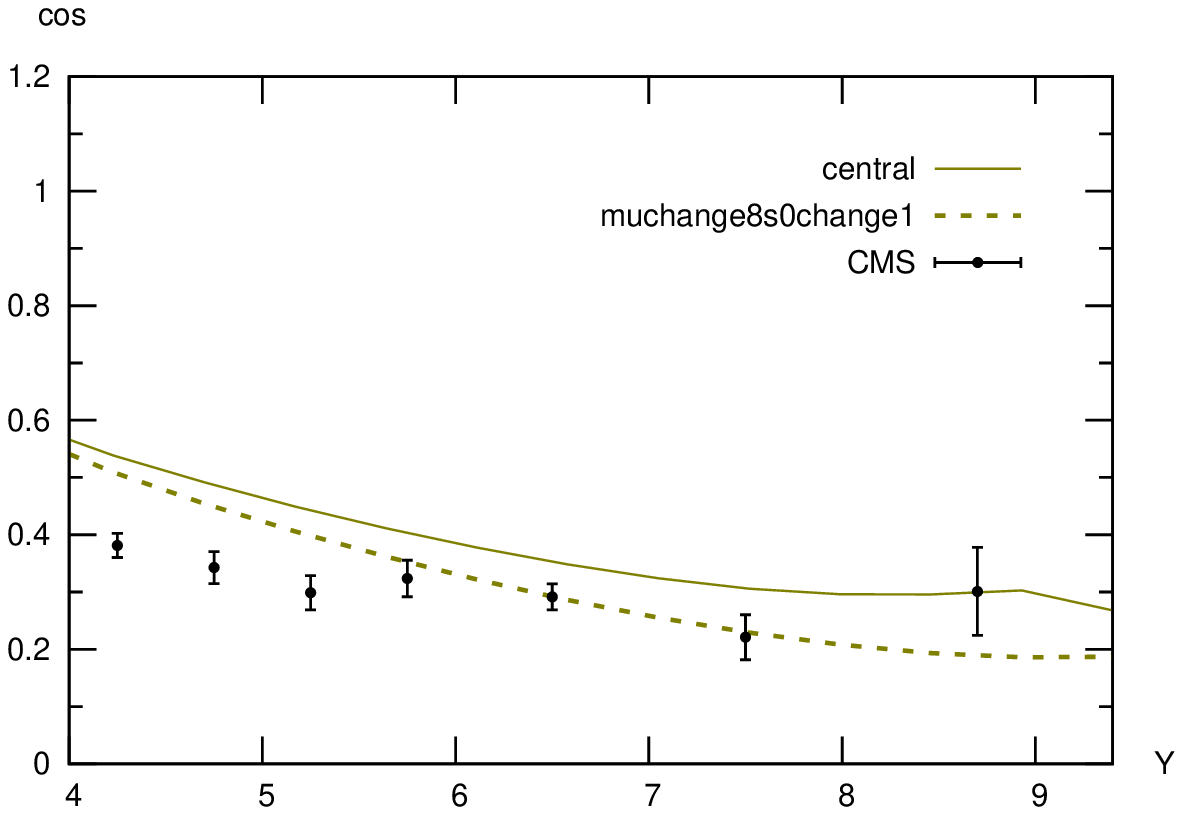}} 
  \end{minipage}
  \begin{minipage}{0.49\textwidth}
    \psfrag{central}[l][l][.65]{\footnotesize $\mu=\sqrt{|\veckjone|\cdot |\veckjtwo|}$}
    \psfrag{muchange8s0change1}[l][l][.65]{\raisebox{-.2cm}{\footnotesize $\mu=8\sqrt{|\veckjone|\cdot |\veckjtwo|}$}}
    \psfrag{CMS}[l][l][.65]{\raisebox{-.3cm}{ \footnotesize CMS data}}
    \psfrag{cos}{\scalebox{0.8}{$\langle \cos 2\varphi\rangle / \langle \cos \varphi\rangle$}}
    \hspace{-.1cm}\scalebox{\scale8}{\includegraphics[width=6.1cm]{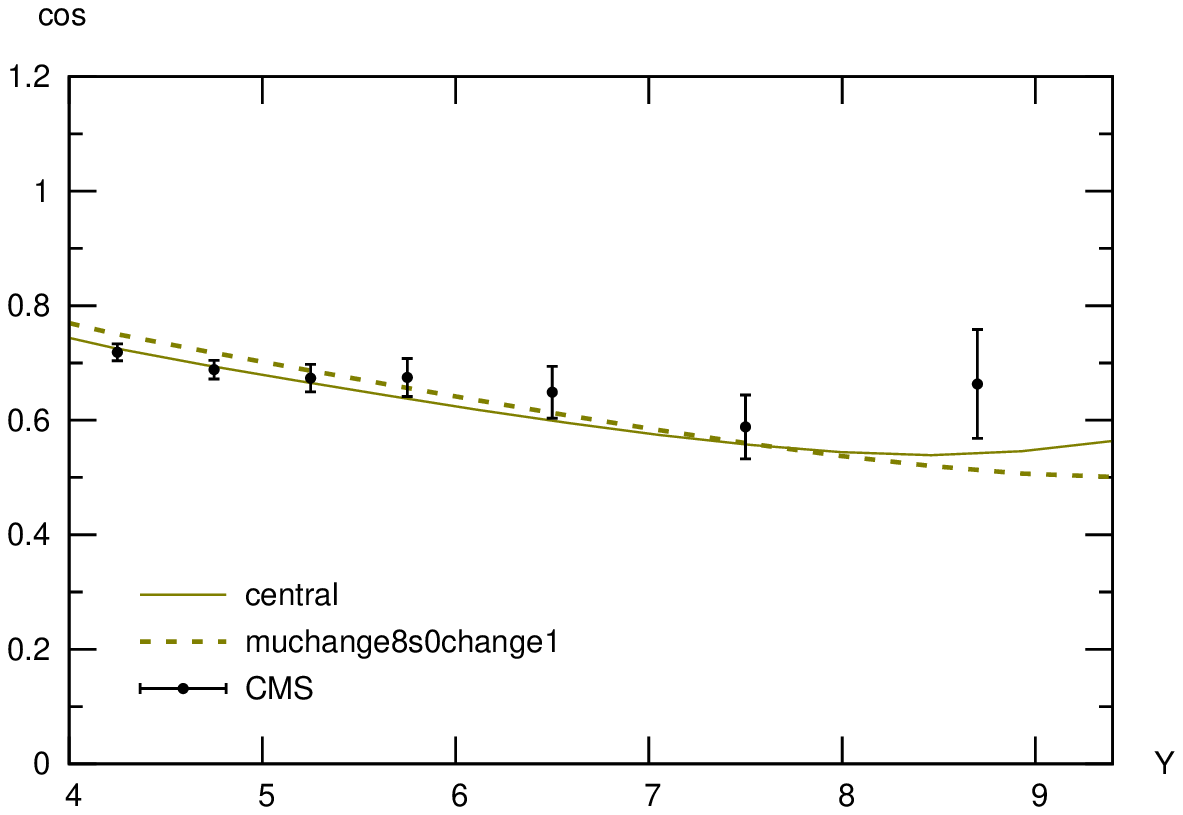}}
  \end{minipage}
  \caption{Comparison of the full NLL BFKL calculation with either a ``natural'' scale $\mu=\sqrt{|\veckjone|\cdot |\veckjtwo|}\,$
  or a large scale $\mu=8\sqrt{|\veckjone|\cdot |\veckjtwo|}\,$ and the CMS data, using symmetric cuts defined in (\protect\ref{sym-cuts}). Left: value of $\avgcosthree$ as a function of the rapidity separation $Y$. Right: value of $\avgcostwo/\avgcos$ as a function of the rapidity separation $Y$.}
  \label{Fig:cos3-cos2cos_scale8}
\end{figure}
%
%%%%%%%%%%%%%%%%   cos3/cos2 and distribution scale 8 muF  %%%%%%
%
%
\begin{figure}[htbp]
  \begin{minipage}{0.49\textwidth}
    \psfrag{central}[l][l][.65]{\footnotesize $\mu=\sqrt{|\veckjone|\cdot |\veckjtwo|}$}
    \psfrag{muchange8s0change1}[l][l][.65]{\raisebox{-.2cm}{\footnotesize $\mu=8\sqrt{|\veckjone|\cdot |\veckjtwo|}$}}
    \psfrag{CMS}[l][l][.65]{\raisebox{-.3cm}{\footnotesize CMS data}}
    \psfrag{cos}{\scalebox{0.8}{$\langle \cos 3\varphi\rangle/\langle \cos 2\varphi\rangle$}}
    \hspace{-.2cm}\scalebox{\scale8}{\includegraphics[width=6.3cm]{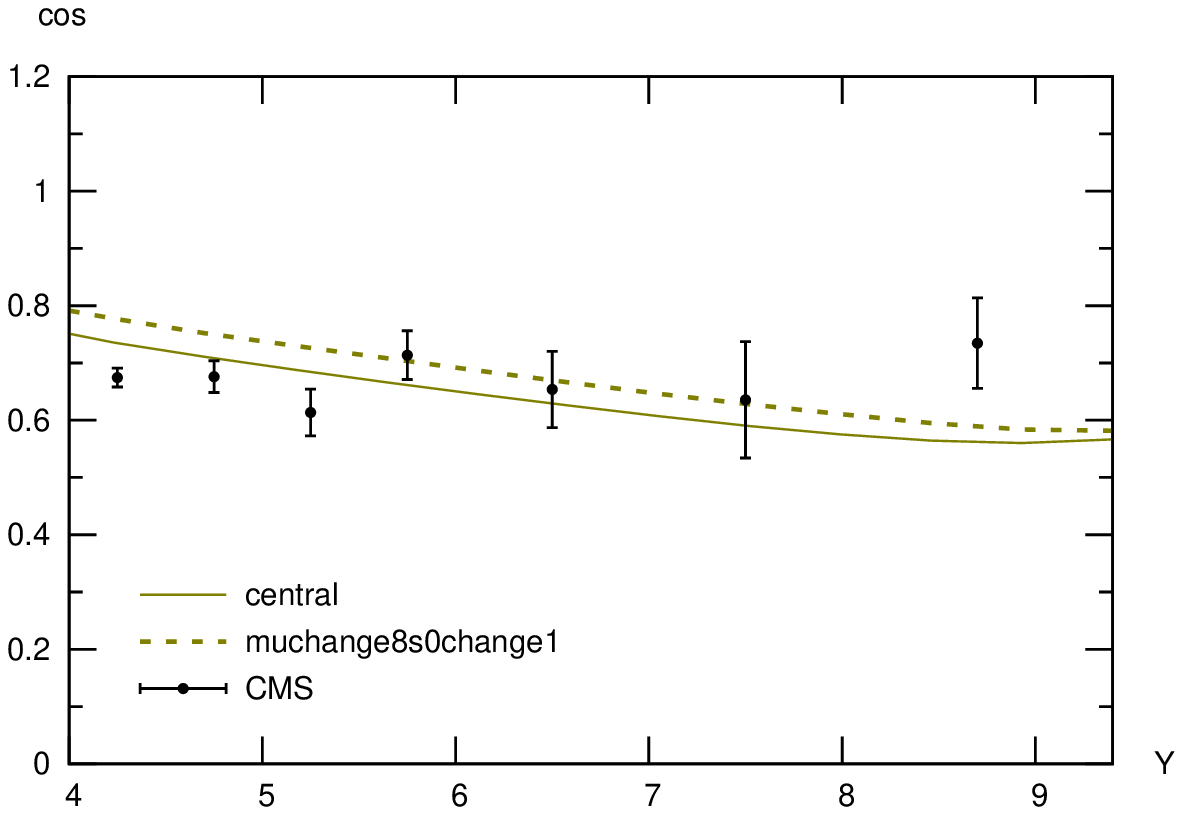}} 
  \end{minipage}
  \begin{minipage}{0.49\textwidth}
    \psfrag{central}[r][r][.65]{\footnotesize $\mu=\sqrt{|\veckjone|\cdot |\veckjtwo|}$}
    \psfrag{muchange8s0change1}[r][r][.65]{\raisebox{-.2cm}{\footnotesize $\mu=8\sqrt{|\veckjone|\cdot |\veckjtwo|}$}}
    \psfrag{CMS}[r][r][.65]{\raisebox{-.3cm}{\footnotesize CMS data}}
    \psfrag{dist}{\raisebox{.1cm}{\scalebox{0.9}{$\frac{1}{\sigma}\frac{d \sigma}{d \varphi}$}}}
    \psfrag{phi}{\raisebox{.1cm}{\scalebox{0.9}{$\varphi$}}}
    \hspace{-1.1cm}\scalebox{.95}{\includegraphics[width=9cm]{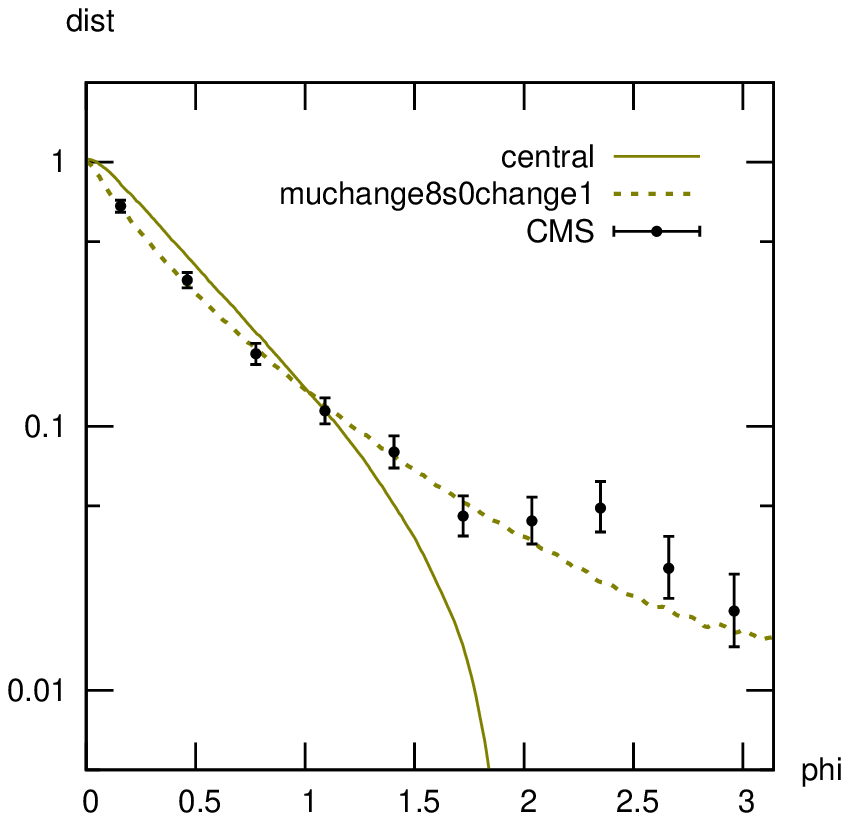}}
  \end{minipage}
  \caption{Comparison of the full NLL BFKL calculation with either a ``natural'' scale $\mu=\sqrt{|\veckjone|\cdot |\veckjtwo|}$
  or a large scale $\mu=8\sqrt{|\veckjone|\cdot |\veckjtwo|}\,$ and the CMS data, using symmetric cuts defined in (\protect\ref{sym-cuts}). Left: value of $\avgcosthree/\avgcostwo$ as a function of the rapidity separation $Y$. Right: value of the azimuthal distribution integrated over the range $6<Y<9.4$ as a function of $\varphi$.}
  \label{Fig:cos3cos2_dist_scale8}
\end{figure}

\section{Conclusions}

We have been able to compare the predictions of our full NLL BFKL calculation of Mueller-Navelet jets with data taken at the LHC thanks to the measurement presented by the CMS collaboration. This comparison shows~\cite{Ducloue:2013tpa} that for the observables $\avgcosn$ a pure LL BFKL treatment or a mixed treatment where the NLL Green's function is used together with LL vertices cannot describe the data. On the other hand, the results of our complete NLL calculation do not agree very well with the data when the scales involved are fixed at their ``natural'' value $\sqrt{|\veckjone|\cdot |\veckjtwo|}$. 
The ratios of these observables are very 
stable with respect to changes of the scales and our calculation describes the data quite well.

The azimuthal distribution has also been measured by CMS, and its description based on our full NLL BFKL treatment with a natural renormalization/factorization scale fails for the nearside configurations. This description, as well as the description of 
$\avgcos,$ $\avgcostwo$ and $\avgcosthree$
is very successful provided one takes a large renormalization/factorization scale, much larger than the natural one. We have shown recently~\cite{Ducloue:2013bva} that this can be understood when fixing the renormalization scale according to the physically 
motivated Brodsky-Lepage-Mackenzie procedure~\cite{Brodsky:1982gc}, in the spirit of Refs.~\cite{Brodsky:1998kn_Brodsky:2002ka,Angioni:2011wj_Hentschinski:2012kr_Hentschinski:2013id}, which indeed leads to a very good description of the CMS data.

To find a fully conclusive evidence for the need of BFKL-type resummation in Mueller-Navelet jets, a comparison with a fixed order (NLO) treatment would be needed.
We have compared our predictions with the ones obtained with the fixed order NLO code \textsc{Dijet} in an asymmetric configuration, as required to get stable results in a fixed order approach.  We found that for the observables $\avgcosn$ no significant difference is observed when taking into account the scale uncertainties. On the other hand, for $\avgcostwo / \avgcos$ and $\avgcosthree / \avgcostwo$ the two calculations lead to noticeably different results. Since these observables are quite stable with respect to scale variations, they are well-suited to study resummation effects at high energy. 
We thus believe that an experimental analysis with different lower cuts on the transverse momenta of the jets
would be of great interest.

\section*{Acknowledgments}

We thank Michel Fontannaz, Cyrille Marquet and Christophe Royon for
providing their codes and for stimulating discussions.
We warmly thank Grzegorz Brona, David d'Enterria, Hannes Jung, Victor Kim and
Maciej Misiura for many discussions and fruitful suggestions on the experimental aspects
of this study.

We thank the organizers, the Tel Aviv French ambassy and the French CEA (IPhT and DSM) for support. 
This work is supported by the Joint Research Activity Study of Strongly Interacting Matter (HadronPhysics3, Grant Agreement n.283286) under the 7th Framework Programm of the European Community and 
by
the Polish Grant NCN No.~DEC-2011/01/B/ST2/03915.

\end{document}